\documentclass[11pt]{article}

\usepackage{amssymb}
\usepackage{amsmath}

\usepackage[pdftex]{graphicx}
\DeclareGraphicsExtensions{.png,.jpg,.eps} 

\usepackage{cases}
\usepackage{color}
\usepackage[ansinew]{inputenc} 
\usepackage{float}

\usepackage{subfigure}
\usepackage{setspace}
\usepackage{natbib}

\newtheorem{Rem}{Remark}

\title{\bf Local Volatility Pricing Models for Long-dated FX Derivatives}

\author{Griselda Deelstra \ and Gr\'{e}gory Ray\'ee \\[3mm]
Department of Mathematics, Universit\'e Libre de Bruxelles,\\
Boulevard du Triomphe, CP 210, Brussels 1050, Belgium,\\
e-mail:  griselda.deelstra@ulb.ac.be \\ grayee@ulb.ac.be
\phantom{..}
 }

\begin{document}

\maketitle

\begin{abstract}
We study the local volatility function in the Foreign Exchange
market where both domestic and foreign interest rates are
stochastic. This model is suitable to price long-dated FX
derivatives. We derive the local volatility function and obtain
several results that can be used for the calibration of this local
volatility on the FX option's market. Then, we study an extension
to obtain a more general volatility model and propose a
calibration method for the local volatility associated to this
model.
\end{abstract}


\section{Introduction}


Recent years, the long-dated FX option's market has grown
considerably. Currently most traded and liquid long-dated FX
Hybrid products are Power-Reverse Dual-Currency swaps (PRDC) (see
for example \citep{Piterbarg2005}) as well as vanilla or exotic
long-dated products such as barrier options. While for short-dated
options (less than 1 year), assuming constant interest rates does
not lead to significant mispricing, for long-dated options the
effect of interest rate volatility becomes increasingly pronounced
with increasing maturity and can become as important as that of
the FX spot volatility. Most of the dealers are using a
three-factor pricing model for long-dated FX products (see
\citep{Piterbarg2005,SippelandOhkoshi}) where the FX spot is
locally governed by a geometric Brownian motion, while each of the
domestic and foreign interest rates follows a Hull-White one
factor Gaussian model \citep{HullandWhite}. Using such a model
does not allow the volatility smile/skew effect encountered in the
FX market to be taken into account, and is therefore not
appropriate to price and hedge long-dated FX products.\\


Different methods exist to incorporate smile/skew effects in the
three-factor pricing model. In the literature, one can find
different approaches which consist either of using a local
volatility for the FX spot or a stochastic volatility and/or jump.
There are many processes that can be used for the stochastic
volatility and their choices will generally depend on their
tractability and solvability. All these models should be
calibrated over the market before being used for pricing. The
calibration is normally based on calculating prices of liquid
products for different strikes and maturity and the parameters of
the model are adjusted until these prices match sufficiently with
the market. However, in most cases it is difficult to derive
analytical formulae, and consequently the calibration procedure
often remains approximative or computationally demanding.
Andreasen suggested in \citep{Andreasen} a stochastic volatility
approach by combining a Heston \citep{heston} stochastic
volatility model with independent stochastic interest rates. He
has derived closed-form Fourier expressions for vanilla options
which are useful for calibration on the vanilla market smile/skew.
He has used an indirect approach in the form of a volatility
displacement parameter to correlate the independent interest rates
with the spot FX rate. In \citep{Antonov}, Antonov et al.\ have
underlined the problem that using non-trivial correlations
destroys the affine structure and exact solvability. By using the
technique of Markovian Projection, they have derived approximation
formulae for the calibration of FX options in a three-factor model
coupled with a Heston stochastic volatility under a full
correlation structure.
In \citep{vanHaastrecht01}, van Haastrecht et al.\ have obtained
the exact pricing of FX options under the three-factor model
coupled with a Sch\"{o}bel and Zhu \citep{SchobelandZhu}
stochastic volatility and the full correlation structure. Their
model could cover Poisson type jump with a trivial extension.
Recently, van Haastrecht and Pelsser considered in
\citep{vanHaastrecht02} the pricing of FX options under
Sch\"{o}bel and Zhu and also the Heston  stochastic volatility
with a multi-factor Gaussian interest rates and a full correlation
structure. More precisely, they derived the characteristic
functions required for the Fourier-based pricing methods under
Sch\"{o}bel and Zhu stochastic volatility. Unfortunately, they did
not obtain a closed-form expression for the characteristic
functions under the Heston stochastic volatility with a full
correlation structure. However, they presented a calibration
method based on the characteristic functions in the special
uncorrelated case by using a projection of the general model onto
the uncorrelated case, or by using it as a control variate for the
general model. Using martingale methods and Fourier inversion
techniques, Ahlip \citep{Ahlip} has derived an analytical formula
for the price of a European call on the spot FX rate in the case
of a stochastic volatility model with stochastic interest rates
where domestic and foreign interest rates are modelled by
Ornstein-Uhlenbeck processes and the instantaneous volatility
follows a mean-reverting Ornstein-Uhlenbeck process correlated
with the spot FX rate. Finally, L.A. Grzelak and C.W. Oosterlee
\citep{Grzelak} have derived semi-closed form approximations for
the forward characteristic function in a foreign exchange model of
Heston-type, in which the domestic and foreign interest rates are
generated by the short-rate process of Hull-White and have
extended the framework by modeling the interest rate by a
stochastic volatility displaced-diffusion Libor
Market Model.\\

A local volatility framework was explored by Piterbarg in
\citep{Piterbarg2005} where the volatility of the spot FX rate was
a function of both the time and the spot itself. He has derived an
approximative formula for the local volatility which allows for a
fast calibration of the model on vanillas. The calibration
essentially captures the ``slope" of the implied
volatility surface but does not exactly fit its convexity.\\

Local volatility models, introduced in 1994 by Dupire
\citep{Dupire} and Derman and Kani \citep{Derman} in the settings
of equity, have the benefit over a stochastic volatility model
that they are Markovian in only one factor since the local
volatility is a deterministic function of both the FX spot and
time. It avoids the problem of working in incomplete markets in
comparison with stochastic volatility models and is therefore more
appropriate for hedging strategies. Local volatility models also
have the advantage to be calibrated on the complete implied
volatility surface, and consequently local volatility models
usually capture more precisely the surface of implied volatilities
than stochastic volatility models. However, a local volatility
model has the drawback that it predicts unrealistic dynamics for
the stock volatility since the volatilities observed in the market
are really stochastic, capable of rising without a movement in
spot FX prices. In \citep{Bossens}, the authors compare
short-dated barrier option market prices with the corresponding
prices derived from either a Dupire local volatility or a Heston
stochastic volatility model both calibrated on the vanilla
smile/skew. It appears from that study that in a simplified world
where exotic option prices are derived either from Dupire local
volatility or from Heston stochastic volatility dynamics, a FX
market characterized by a mild skew (USDCHF) exhibits mainly a
stochastic volatility behavior, and that FX markets characterized
by a dominantly skewed implied volatility (USDJPY) exhibit a
stronger local volatility component. This observation also
underlines that calibrating a stochastic model to the vanilla
market is by no means a guarantee that exotic options will be
priced correctly \citep{Schoutens}, as the vanilla market carries
no information about the smile dynamics. The market dynamics could
be better approximated by a hybrid volatility model that contains
both stochastic volatility dynamics and local volatility ones.
This approach has to our knowledge never been studied in a
stochastic interest rates framework but gives positive results for
short dated options when interest rates are assumed to be constant
(see for example \citep{lipton2,lipton3,Madan,Tavella}).
In the constant interest rates case, once the local volatility
surface is available, the new mixed volatility can be computed by
multiplying this local volatility with a ratio of integrals that
depend on the joint density of the FX spot and the stochastic
volatility. This density can be determined by numerically solving
the associated two-dimensional Kolmogorov forward PDE.
\\

The study of the local volatility and its calibration in a
three-factor model with local volatility can therefore be
motivated by hedging arguments but is also considerably useful for
the calibration of hybrid volatility models. In this paper we
derive the local volatility function in a three-factor model with
local volatility where we have three sources of randomness: the FX
spot with a local volatility and the domestic/foreign interest
rates.\\

In a one-factor Gaussian model, the local volatility surface is
generally built by using the Dupire's formula where partial
derivatives of call options with respect to strikes and maturities
are calculated by finite differences and where the real implied
volatility surface is an interpolation of a finite set of market
call prices. In a three-factor framework with local volatility,
the expression of the local volatility becomes more complicated as
it also depends on a particularly complicated expectation
where no closed form expression exists and it is not directly
related to European call prices or other liquid products. Its
calculation can obviously be done by numerical integration methods
or Monte Carlo simulations. To enable realizations of the
numerical integrations you need the forward probability
distribution of the spot FX rate and the domestic and foreign
interest rates up to maturity which is a solution of a three-dimensional forward PDE.\\

An alternative approach is to calibrate the local volatility from
stochastic volatility models by establishing links between local
and stochastic volatility.
Extracting the local volatility surface from a stochastic
volatility model rather than by using the market implied
volatility surface presents several advantages. First, the market
implied volatility surface can in practice only be an
interpolation of a finite set of available market data. As a
consequence a local volatility surface built from an approximative
implied volatility surface is often unstable. Stochastic
volatility models can be calibrated by using fast algorithms like
Fast Fourier Transforms (FFT) (see for example \citep{Ahlip,
Andreasen, vanHaastrecht01}) and the local volatility surface
extracted from the calibrated stochastic volatility model is
really smooth. In this paper we will present some mimicking
properties that links the three-factor model with a local
volatility to the same model with a stochastic volatility rather
than a local volatility. These properties will allow us to obtain
explicit expressions to
construct the local volatility surface.\\

Finally, we derive a link between the three-factor model with
local volatility and a hybrid volatility model where the
volatility of the spot FX rate mixes a stochastic volatility with
a local volatility. Knowing the local volatility function
associated to the three-factor model with local volatility, we
propose a calibration method for the local
volatility in the four-factor hybrid volatility model.\\

Our objective is to expose theoretical results about the local
volatility function and its calibration. Numerical implementation
of the model to the long-dated FX options market will be studied in a subsequent paper.\\

This paper is organized as follows: We begin by defining the
three-factor model with local volatility in section \ref{section :
the model}. Then, in section  \ref{section : Dupire's like formula
} we derive the local volatility expression for this model. First,
we derive from the three-dimensional Fokker-Plank equation for the
forward probability density function, a ``simpler" one-dimensional
PDE. This PDE is used in the derivation of the local volatility
function by differentiating European call price expressions with
respect to the strike and the maturity. Section \ref{section : loc
vol calibration} is devoted to the calibration of this local
volatility function. In section \ref{calibraiation numerical
approaches}, we present two numerical approaches based on
respectively Monte Carlo simulations and PDE numerical resolution.
In section \ref{section: calibration Comparison between local
volatility} we obtain a link between the local volatility function
derived in a three-factor framework and the one coming from the
simple one-factor Gaussian model. Next, in section \ref{section :
mimicking stoch vol}, we derive a link between the three-factor
model with a stochastic volatility for the spot FX rate and the
one where the spot FX rate volatility is a local volatility. This
link provides a relationship between our local volatility function
and future instantaneous spot FX rate volatilities. Finally, in
section \ref{section:Extension}, we study an extension of the
three-factor model with local volatility. We derive a link between
the local volatility function associated to the three-factor model
and the local volatility function present in the four-factor
hybrid volatility model and propose a calibration procedure for
the local volatility function associated to this four-factor
hybrid volatility model. Conclusions are presented in section
\ref{section : Conclusion}.


\section{The three-factor pricing model with local volatility \label{section : the model}}

In this paper we consider the three-factor model where the spot FX
rate volatility is a deterministic function of both the time and
the FX spot itself. This function is known as `local volatility'.
In this model, the spot FX rate $S$ is governed by the following
dynamics

\begin{eqnarray}
dS(t) = (r_{d}(t) - r_{f}(t))  S(t)  dt + \sigma(t,S(t)) S(t)
dW^{DRN}_{S}(t), \label{FXspot_dynamic}
\end{eqnarray}

\noindent where the volatility of the spot FX rate is a local
volatility $\sigma(t,S(t))$ and where domestic and foreign
interest rates denoted by $r_d$ and $r_f$ respectively follow a
Hull-White one factor Gaussian model \citep{HullandWhite} defined
by the Ornstein-Uhlenbeck processes

\begin{numcases}{}
dr_{d}(t) = [\theta_{d}(t) - \alpha_{d}(t)  r_{d}(t) ]  dt +
\sigma_{d}(t) dW^{DRN}_{d}(t), & \label{rd_HW_dynamic} \\
dr_{f}(t) = [\theta_{f}(t) - \alpha_{f}(t)  r_{f}(t) - \rho_{fS}
\sigma_{f}(t) \sigma(t,S(t)) ]  dt + \sigma_{f}(t)
dW^{DRN}_{f}(t), & \label{rf_HW_dynamic}
\end{numcases}

\noindent where $\theta_{d}(t), \alpha_{d}(t), \sigma_{d}(t),
\theta_{f}(t), \alpha_{f}(t), \sigma_{f}(t)$ are deterministic
functions of time. Equations (\ref{FXspot_dynamic}),
(\ref{rd_HW_dynamic}) and (\ref{rf_HW_dynamic}) are expressed in
the domestic risk-neutral measure (DRN).\\


Foreign and domestic zero-coupon bonds defined by
$P_i(t,T)=\mathbf{E}^{Q_i}[ e^{- \int_{t}^{T} r_{i}(s)ds} \mid
\mathcal{F}_t]$, $i=d,f$ satisfy the evolution

\begin{numcases}{}
dP_{d}(t,T) = r_{d}(t)  P_{d}(t,T)  dt - \sigma_{d}(t,T)
P_{d}(t,T) dW^{DRN}_{d}(t), & \label{eqution 1.2} \\
dP_{f}(t,T) = [r_{f}(t) + \rho_{fS} \sigma_{f}(t,T)
\sigma(t,S(t))] P_{f}(t,T)  dt - \sigma_{f}(t,T)  P_{f}(t,T)
dW^{DRN}_{f}(t),\label{eqution 1.3} &
\end{numcases}

\noindent where their volatility structures are given by

\begin{eqnarray}
\sigma_{i}(t,T) &=& \sigma_{i}(t) \int_{t}^{T} e^{ - \int_{t}^{s}
\alpha_{i}(u) du} ds, \ \ i=d,f. \nonumber
\end{eqnarray}

More details about stochastic interest rates in the Hull-White one
factor Gaussian model and explicit expression for the price of
zero-coupon bonds under this framework can be found in
\citep{Brigo-Mercurio}.

Finally, in this paper we assume that
$(W^{DRN}_{S}(t),W^{DRN}_{d}(t),W^{DRN}_{f}(t))$  is a Brownian
motion under the domestic risk-neutral measure $Q_{d}$ with
correlation matrix

\[
\begin{pmatrix}
1 & \rho_{Sd} & \rho_{Sf} \\
\rho_{Sd} & 1 & \rho_{df} \\
\rho_{Sf} & \rho_{df} & 1
\end{pmatrix}.
\]

\section{The local volatility function \label{section : Dupire's like formula } }

In this section we derive the expression of the local volatility
function by using the same technique as Dupire \citep{Dupire} and
Derman and Kani \citep{Derman} which consists in differentiating
European call price expressions with respect to the strike and the
maturity.

\subsection{Forward PDE \label{section : fwd PDE}}

Consider a world where both the spot FX rate as well as the
domestic and the foreign interest rates are stochastic. Under the
assumption of absence of arbitrage opportunities, and working
under the $t$-forward measure $Q_t$ (where the domestic
zero-coupon bond is the numeraire), the present value
$V(S(0),r_d(0),r_f(0),t=0)$ of a derivative that pays off
$V(S(t),r_d(t),r_f(t),t)$ at time $t$ is given by

\begin{eqnarray}
V(S(0),r_d(0),r_f(0),t=0) &=& P_d(0,t) \mathbf{E}^{Q_{t}}[
V(S(t),r_d(t),r_f(t),t)]  \nonumber \\
&=& P_d(0,t) \int \int \int  V(x,y,z,t) \phi_{F}(x,y,z,t) dx dy dz
\nonumber \label{present value QTF}
\end{eqnarray}

\noindent where $\phi_{F}(x,y,z,t)$ corresponds to the $t$-forward
probability density.\\

Following the same approach as in `Equity Hybrid Derivatives'
\citep{Overhaus} (readapting the method in the context of FX
market), we can get the following Fokker-Plank equation for
$\phi_{F}(x,y,z,t)$

\begin{eqnarray}
0 &=& \frac{\partial \phi_{F}}{\partial t} + (r_{d}(t) - f_d(0,t))
\ \phi_{F} +  \frac{\partial [(r_{d}(t) - r_{f}(t)) S(t) \phi_{F}]
}{\partial x } + \frac{\partial [(\theta_{d}(t) -
\alpha_{d}(t) \ r_{d}(t) ) \phi_{F}] }{\partial y }  \nonumber \\
& & + \frac{\partial [(\theta_{f}(t) - \alpha_{f}(t) \ r_{f}(t) -
\rho_{fS} \sigma_{f}(t) \sigma(t,S(t)) ) \phi_{F}] }{\partial z }
- \frac{1}{2} \frac{\partial^{2} [\sigma^{2}(t,S(t)) S^2(t)
\phi_{F}]}{\partial x^{2} } - \frac{1}{2} \frac{\partial^{2}
[\sigma^{2}_{d}(t) \phi_{F}] }{\partial y^{2} }  \nonumber \\
& & - \frac{1}{2} \frac{\partial^{2} [\sigma^{2}_{f}(t) \phi_{F}]
}{\partial z^{2} } - \frac{\partial^{2} [\sigma(t,S(t)) S(t)
\sigma_{d}(t) \rho_{Sd} \phi_{F}] }{\partial x \partial y } -
\frac{\partial^{2} [\sigma(t,S(t)) S(t) \sigma_{f}(t) \rho_{Sf}
\phi_{F} ]}{\partial x
\partial z } - \frac{\partial^{2} [\sigma_{d}(t) \sigma_{f}(t) \rho_{df}
\phi_{F}]}{\partial y \partial z }. \nonumber \\
\label{PDE phiTF}
\end{eqnarray}

Equation (\ref{PDE phiTF}) is a forward PDE since it is solved
forward in time with the initial condition at time $t=0$ given by
$\phi_{F}(x,y,z,t)=\delta(x-x_0,y-y_0,z-z_0)$, where $\delta$ is
the Dirac delta function and $x_0$, $y_0$ and $z_0$ correspond to
the values at time $t=0$ of the spot FX rate, the domestic and
foreign interest rates respectively. This initial condition means
that at time $t=0$ we are sure that the spot FX rate $S(0)$ equals
$x_0$, the domestic interest rate $r_d(0)$ equals $y_0$ and the
foreign interest
rate $r_f(0)$ equals $z_0$.\\



From equation (\ref{PDE phiTF}) we will derive a ``simpler"
one-dimensional PDE only involving the dependence in $x$ of the
forward density function. This equation will be useful in the
derivation of the local volatility expression in section
\ref{section: loc vol derivation Dupire method}.

Let us denote by $q_{F}(x,z,t)$ the function defined as the
integral of $\phi_{F}(x,y,z,t)$ over the whole range of $y$

\begin{eqnarray}
q_{F}(x,z,t)=\int_{-\infty}^{+\infty} \phi_{F}(x,y,z,t) dy.
\label{qTF}
\end{eqnarray}

Making the realistic assumptions that $ \underset{y \rightarrow
\pm \infty} \lim \phi_{F}(x,y,z,t) = 0 $ and that partial
derivatives of $\phi_{F}$ with respect to $x, y$ and $z$ tend also
to zero when $y$ tends to infinity, we obtain, after integrating
(\ref{PDE phiTF}) with respect to $y$, the following two
dimensional PDE

\begin{eqnarray}
0 &=& \frac{\partial q_{F}}{\partial t} + \int (r_{d}(t) -
f_d(0,t)) \phi_{F} dy + \frac{\partial }{\partial x } \left( \int
(r_{d}(t) - r_{f}(t)) S(t) \phi_{F} dy \right) \nonumber \\
& & + \frac{\partial [(\theta_{f}(t) - \alpha_{f}(t) \ r_{f}(t) -
\rho_{fS} \sigma_{f}(t) \sigma(t,S(t)) ) q_{F}] }{\partial z }  -
\frac{1}{2} \frac{\partial^{2}
[\sigma^{2}(t,S(t)) S^2(t) q_{F}]}{\partial x^{2} }   \nonumber \\
& &- \frac{1}{2} \frac{\partial^{2} \sigma^{2}_{f}(t) q_{F}
}{\partial z^{2} } - \frac{\partial^{2} \sigma(t,S(t)) S(t)
\sigma_{f}(t) \rho_{Sf} q_{F} }{\partial x
\partial z }. \label{PDE qTF}
\end{eqnarray}

Finally, we define $p_{F}(x,t)$ as the integral of $q_{F}(x,z,t)$
over the whole range of $z$

\begin{eqnarray}p_{F}(x,t) = \int_{-\infty}^{+\infty} q_{F}(x,z,t) dz =
\int_{-\infty}^{+\infty} \int_{-\infty}^{+\infty}
\phi_{F}(x,y,z,t) dy dz.
\end{eqnarray}

Assuming that $ \underset{z \rightarrow \pm \infty} \lim
q_{F}(x,z,t) = 0 $ and that partial derivatives of $q_{F}$ with
respect to $x$ and $z$ tend also to zero when $z$ tends to the
infinity, then integrating (\ref{PDE qTF}) with respect to $z$
reduces the PDE (\ref{PDE qTF}) into the following one-dimensional
 PDE

\begin{eqnarray}
0 &=& \frac{\partial p_{F}}{\partial t} + \int \int (r_{d}(t) -
f_d(0,t)) \phi_{F} dy dz + \frac{\partial }{\partial x } \left(
\int
\int (r_{d}(t) - r_{f}(t)) S(t) \phi_{F} dy dz \right)  \nonumber \\
& & - \frac{1}{2} \frac{\partial^{2} [\sigma^{2}(t,S(t)) S^{2}(t)
p_{F}]}{\partial x^{2} }. \label{PDE pTF}
\end{eqnarray}

\subsection{The local volatility derivation \label{section: loc vol derivation Dupire method}}

In this section we outline how to derive the expression of the
local volatility function associated to the three-factor model
with local volatility by differentiating the expression of a
European call price ($C(K,t)$) with respect to its strike $K$ and
its maturity $t$. In order to simplify our calculations, we will
work with the forward call price $\widetilde{C}(K,t)$ defined by

\begin{eqnarray}
\widetilde{C}(K,t)=\frac{C(K,t)}{P_d(0,t)} = \mathbf{E}^{Q_t} [
(S(t)-K)^{+}]
 = \int \int \int_{K}^{+ \infty} (x-K) \phi_{F}(x,y,z,t) dx dy dz.
 \label{Ctild}
\end{eqnarray}

Differentiating equation (\ref{Ctild}) twice with respect to $K$,
we obtain

\begin{eqnarray}
\frac{\partial \widetilde{C}(K,t)}{\partial K} &=& \int \int
\int_{K}^{+ \infty} - \phi_{F}(x,y,z,t) dx dy dz = -
\mathbf{E}^{Q_t}[
\textbf{1}_{ \{S(t)>K\}} ], \label{dC1} \\
\frac{\partial^{2} \widetilde{C}(K,t)}{\partial K^{2}} &=& \int
\int \phi_{F}(K,y,z,t) dy dz \equiv p_{F}(K,t). \label{dC2}
\end{eqnarray}

Differentiating equation (\ref{Ctild}) with respect to the
maturity and using equation (\ref{PDE pTF}) leads to

\begin{eqnarray}
\frac{\partial \widetilde{C}(K,t)}{\partial t} &=& \int \int
\int_{K}^{+ \infty} (S(t)-K) \frac{\partial
\phi_{F}(x,y,z,t)}{\partial t} dx dy dz
\nonumber \\
&=& \int_{K}^{+ \infty} (S(t)-K) \frac{\partial p_{F}}{\partial t}
dx
\nonumber \\
&=& \int_{K}^{+ \infty} (S(t)-K) \{- \int \int (r_{d}(t) -
f_d(0,t)) \phi_{F} dy dz   \nonumber \\
& & - \frac{\partial }{\partial x } ( \int \int (r_{d}(t) -
r_{f}(t)) S(t) \phi_{F} dy dz) + \frac{1}{2} \frac{\partial^{2}
[\sigma^{2}(t,S(t)) S^{2}(t) p_{F}]}{\partial x^{2} } \}dx .
\nonumber
\end{eqnarray}

Integrating by parts several times and using equation (\ref{dC2})
we get

\begin{eqnarray}
\frac{\partial \widetilde{C}(K,t)}{\partial t} &=& f_d(0,t)
\widetilde{C}(K,t) + \int \int \int_{K}^{+ \infty} [r_{d}(t) K -
r_{f}(t) S(t) ] \phi_{F}(x,y,z,t) dx dy dz \nonumber \\
& & + \frac{1}{2}(\sigma(t,K) K)^2 \frac{\partial^{2}
\widetilde{C}(K,t)}{\partial K^{2}} \nonumber \\
& = & f_d(0,t) \widetilde{C}(K,t) + \mathbf{E}^{Q_t}[( r_{d}(t) K
-
r_{f}(t) S(t)) \textbf{1}_{ \{S(t)>K\} }\ ] \nonumber \\
& &  + \frac{1}{2}(\sigma(t,K) K)^2 \frac{\partial^{2}
\widetilde{C}(K,t)}{\partial K^{2}}.
\end{eqnarray}

This leads to the following expression for the local volatility
surface in terms of the forward call prices $\widetilde{C}(K,t)$

\begin{eqnarray}
\sigma^{2}(t,K) = \frac{\frac{\partial
\widetilde{C}(K,t)}{\partial t} -f_d(0,t) \widetilde{C}(K,t) -
\mathbf{E}^{Q_t}[( r_{d}(t) K - r_{f}(t) \ S(t)) \textbf{1}_{
\{S(t)>K\} }\ ] }{\frac{1}{2} K^2 \frac{\partial^{2}
\widetilde{C}(K,t)}{\partial K^{2}}}. \label{locvol Ctild}
\end{eqnarray}

The (partial) derivatives of the forward call price with respect
to the maturity and the strike can be easily calculated

\begin{eqnarray}
\frac{\partial \widetilde{C}(K,t)}{\partial t} &=& \frac{\partial
[\frac{C(K,t)}{P_{d}(0,t)}]}{\partial t} =  \frac{\partial
C(K,t)}{\partial t} \frac{1}{P_{d}(0,t)} + f_{d}(0,t)
\widetilde{C}(K,t),  \label{dctild dt}  \\
\frac{\partial^{2} \widetilde{C}(K,t)}{\partial K^{2}}  & = &
\frac{\partial^{2} [\frac{C(K,t)}{P_{d}(0,t)}]}{\partial K^{2}} =
\frac{1}{P_{d}(0,t)} \frac{\partial^{2} C(K,t)}{\partial K^{2}}.
\label{dctild dK2}
\end{eqnarray}

Substituting these expressions (\ref{dctild dt}) and (\ref{dctild
dK2}) into (\ref{locvol Ctild}), we obtain the expression of the
local volatility $\sigma^{2}(t,K)$ in terms of call prices
$C(K,t)$

\begin{eqnarray}
\sigma^{2}(t,K) = \frac{\frac{\partial C(K,t)}{\partial t} -
P_{d}(0,t) \mathbf{E}^{Q_t}[( r_{d}(t) K - r_{f}(t) \ S(t))
\textbf{1}_{ \{S(t)>K\} }\ ]
  }{ \frac{1}{2} K^2 \frac{\partial^{2} C(K,t)}{\partial
 K^{2}}}. \label{locvol C}
\end{eqnarray}

Unfortunately, this extension of the Dupire formula is not easily
applicable for calibration over the market since there seems no
immediate way to link the expectation term with European option
prices or other liquid products. However, we present in section
\ref{section : loc vol calibration}
four different methods to calibrate this local volatility function.\\

Finally, we underline the fact that when assuming deterministic
interest rates, equation (\ref{locvol C}) reduces to the simple
Dupire formula corresponding to the one factor Gaussian case. This
formula can easily be derived by a similar reasoning as above but
in a one factor framework.

\begin{eqnarray}
\sigma^{2}(t,K) = \frac{ \frac{\partial C(K,t)}{\partial t} +
(r_{d}(t) - r_{f}(t)) K \frac{\partial C(K,t)}{\partial K}  +
r_{f}(t) C(K,t)}{\frac{1}{2}K^2 \frac{\partial^{2}
C(K,t)}{\partial K^{2}}}. \label{detreministic locvol C}
\end{eqnarray}

\begin{Rem}
The market often quotes options in terms of implied volatilities
$\sigma_{imp}$ instead of option prices. Consequently, it is more
convenient to express the local volatility in terms of implied
volatilities than option prices. As the implied volatility  of an
option with price $C(K,T)$ is defined through the Black-Scholes
formula ($C^{mkt}(K,T) = C^{BS}(K,T,\sigma_{imp})$), the
derivatives of call prices in equation (\ref{detreministic locvol
C}) can be computed through the chain rule, and this leads to the
following equation (see \citep{Wilmott2})

\begin{eqnarray}
\sigma^{2}(t,K) = \frac{ \sigma^2_{imp} + 2t \sigma_{imp}
\frac{\partial \sigma_{imp}}{\partial t} + 2(r_{d}(0) - r_{f}(0))
K t \sigma_{imp} \frac{\partial \sigma_{imp}}{\partial K} }{(1+ K
d_+ \sqrt{t} \frac{\partial \sigma_{imp}}{\partial K})^2 + K^2 t
\sigma_{imp} ( \frac{\partial^2 \sigma_{imp}}{\partial K^2}-d_+
(\frac{\partial \sigma_{imp}}{\partial K})^2 \sqrt{t})}.
\label{det_implied_locvol C}
\end{eqnarray}

Using the same approach, the local volatility expression
(\ref{locvol C}) can be written in terms of implied volatilities
$\sigma_{imp}$,

\begin{eqnarray}
\sigma^{2}(t,K) = \sigma_{imp} \frac{e^{-r_f(0) t} S(0)
\{\mathcal{N}'(d_+) (\sigma_{imp}+2 t \frac{\partial
\sigma_{imp}}{\partial t}) - 2\sqrt{t} r_f(0) \mathcal{N}(d_+)\} +
2\sqrt{t} \{ r_d(0) K e^{-r_d(0) t} \mathcal{N}(d_-) + E \} }{
e^{-r_f(0) t} S(0) \mathcal{N}'(d_+) \{ (1+ K d_+ \sqrt{t}
\frac{\partial \sigma_{imp}}{\partial K})^2 + K^2 t \sigma_{imp} (
\frac{\partial^2 \sigma_{imp}}{\partial K^2}-d_+ (\frac{\partial
\sigma_{imp}}{\partial K})^2 \sqrt{t}) \}} \nonumber
\\ \label{locvol_implied C}
\end{eqnarray}

 where

\begin{eqnarray}
 E = P_{d}(0,t) \mathbf{E}^{Q_t}[( r_{d}(t) K - r_{f}(t) \ S(t)) \textbf{1}_{ \{S(t)>K\} }], \nonumber
\end{eqnarray}

\begin{eqnarray}
d_{\pm} = \frac{  log \frac{S(0)}{K} + (r_{d}(0) - r_{f}(0) \pm
\frac{\sigma^2_{imp}}{2} ) t}{\sigma_{imp} \sqrt{t}}, \nonumber
\end{eqnarray}

\begin{eqnarray}
\mathcal{N}(x) = \int_{-\infty}^{x} \frac{1}{\sqrt{2 \pi}}
e^{\frac{-z^2}{2}} dz, \nonumber
\end{eqnarray}

\begin{eqnarray}
\mathcal{N}'(x) = \frac{1}{\sqrt{2 \pi}} e^{\frac{-x^2}{2}}.
\nonumber
\end{eqnarray}

\end{Rem}


\section{Calibrating the Local Volatility \label{section : loc vol calibration}}

Before using a model to price any derivatives, it is usual to
calibrate it on the vanilla market which means that you should be
able to price vanilla options with your model such that the
resulting implied volatilities match the market-quoted ones. More
precisely you have to determine all parameters present in the
different stochastic processes which define the model in such a
way that all European option prices derived in the model are as
consistent as possible with
the corresponding market ones.\\

The calibration procedure for the three-factor model with local
volatility can be decomposed in three steps: (i) Parameters
present in the Hull-White one-factor dynamics for the domestic and
foreign interest rates, $\theta_{d}(t), \alpha_{d}(t),
\sigma_{d}(t), \theta_{f}(t), \alpha_{f}(t), \sigma_{f}(t)$, are
chosen to match European swaption / cap-floors values in their
respective currencies. Methods for doing so are well developed in
the literature (see for example \citep{Brigo-Mercurio}). (ii) The
three correlation coefficients of the model, $\rho_{Sd},
\rho_{Sf}$ and $\rho_{df}$ are usually estimated from historical
data. (iii) After these two steps, the calibration problem
consists in finding the local volatility function of the spot FX
rate which is consistent with an implied volatility surface. In
\citep{Piterbarg2005}, Piterbarg derives an approximative formula
for European call prices in the three-factor model where the local
volatility function for the spot FX rate is a parametric function
of the form $\sigma(t,S(t))= \nu(t)
(\frac{S(t)}{L(t)})^{\beta(t)-1} $, where $\nu(t)$ is the relative
volatility function, $\beta(t)$ is a time-dependent elasticity of
variance (CEV) and $L(t)$ is a time-dependent scaling constant.
The calibration procedure of Piterbarg consists in determining the
functions $\nu(t)$ and $\beta(t)$ such that when pricing any
European call with his approximative call valuation formula, the
price he gets is as close as possible to the market call price.
Next sections will be devoted to other calibration methods for the
local volatility function $\sigma(t,S(t))$ based
on the exact expression of this function in a three-factor context.\\

\subsection{Numerical approaches \label{calibraiation numerical approaches}}

\subsubsection{A Monte Carlo approach}

In this section we present a first calibration method for the
local volatility expression (\ref{locvol C}) derived in section
\ref{section : Dupire's like formula }. In this approach, the
expectation, $\mathbf{E}^{Q_T}[( r_{d}(T) K - r_{f}(T)S(T))
\textbf{1}_{ \{S(T)>K\} }\ ]$ is approximated by using Monte Carlo
simulations up to a fixed time $t=T$. To calculate numerically
this expectation we have to simulate the FX spot rate $S(t)$ and
both the domestic and foreign interest rates up to time $T$
starting from the initial market prices $S(0)$, $r_d(0)$ and
$r_f(0)$ respectively. Since the expectation is expressed under
the measure $Q_T$, we have to use the dynamics of $S(t)$, $r_d(t)$
and $r_f(t)$ under this last measure,

\begin{numcases}{}
dS(t) = [r_{d}(t) - r_{f}(t) - \sigma(t,S(t))  \sigma_{d}(t) b_d(t,T)  \rho_{Sd}]  S(t)  dt + \sigma(t,S(t))  S(t) dW^{TF}_{S}(t), & \nonumber \\
dr_{d}(t) =  [\theta_d(t) - \alpha_{d}  r_d(t) - \sigma^2_{d}(t)
b_d(t,T)]
dt + \sigma_{d}(t) dW^{TF}_{d}(t), & \nonumber \\
dr_{f}(t) = [\theta_{f}(t) - \alpha_{f}  r_{f}(t) - \rho_{fS}
\sigma_{f}(t) b_f(t,T) \sigma(t,S(t))- \sigma_{d}(t) b_d(t,T)
\sigma_{f}(t) b_f(t,T) ] \rho_{df}  dt + \sigma_{f}(t)
dW^{TF}_{f}(t) . \nonumber &
\end{numcases}

\noindent where $b_j(t,T)= \frac{1}{\alpha_{j}}
(1-e^{-\alpha_{j}(T-t)})$,
$j=d,f$ (assuming that $\alpha_{j}$ are constant). \\

The idea of the Monte Carlo method is to simulate $n$ times (i.e.
$n$ scenarios) the stochastic variables $S(t)$, $r_d(t)$ and
$r_f(t)$ up to time $T$, by using for example Euler
discretisations. The expectation is approximated by:

\begin{eqnarray}
\mathbf{E}^{Q_T}[ (r_{d}(T) K - r_{f}(T)S(T)) \textbf{1}_{
\{S(T)>K\} } ] \cong \frac{1}{n} \sum_{i=1}^{n} (r^i_d(T) K -
r^i_{f}(T) S^i(T)) \textbf{1}_{ \{S^i(T)>K\}}
\end{eqnarray}

\noindent where $i$ corresponds to the $i^{th}$-scenario $i=1,...,n$.\\

As we have to know the local volatility function up to time $T$ to
simulate the path for $S(t)$ and $r_f(t)$, the only way to work is
forward in time. To begin, we have to determine the local
volatility function at the first time step $T=T_1$ for all strike
$K$. At this first step we assume that the initial local
volatility is equal to the deterministic local volatility given by
equation (\ref{detreministic locvol C}). Note that this local
volatility is directly obtained by using market data (see equation
(\ref{det_implied_locvol C})). More precisely, by this choice, we
assume that for a ``small time period", interest rates are
constant and in this case, the local volatility expression
(\ref{locvol C}) reduces to (\ref{detreministic locvol C}).
Knowing that local volatility function we can simulate $S(T_1)$
and $r(T_1)$. Then we can compute the expectation
$\mathbf{E}^{Q_{T_1}}[ (r_{d}(T_1) K - r_{f}(T_1)S(T_1))
\textbf{1}_{\{S(T_1)>K\} }]$ for all $K$ by using:

\begin{eqnarray}
\mathbf{E}^{Q_{T_1}}[ (r_{d}(T_1) K - r_{f}(T_1)S(T_1))
\textbf{1}_{ \{S(T_1)>K\} } ] \cong \frac{1}{n} \sum_{i=1}^{n}
(r^i_d(T_1) K - r^i_{f}(T_1) S^i(T_1)) \textbf{1}_{
\{S^i(T_1)>K\}}
\end{eqnarray}

This allows us to get the local volatility expression at time
$T_1$, $\sigma^{2}(T_1,K)$, for all strike $K$.

Following the same procedure we can easily calibrate the local
volatility at time $T_2$ by using the local volatility obtained at
time $T_1$ and also the simulated path until time $T_1$. Following
this procedure we are able to generate the local volatility
expression up to a final date $T=T_k$.

\subsubsection{A PDE approach}

The strategy is to solve the forward equation (\ref{PDE phiTF})
forwards one step at a time, starting with a local volatility
$\sigma(0,S(0))$ at time $t_0=0$. At the first time step
$t_1=t_0+\Delta t$, we can generate the forward joint transition
densities $ \phi_F(x,y,z,t_0+\Delta t) $ by solving the forward
PDE using the initial condition
$\phi_{F}(x,y,z,t_0)=\delta(x-x_0,y-y_0,z-z_0)$ at time $t_0=0$.
Knowing the forward joint transition densities we can calculate
the expectation in (\ref{locvol C}) namely $\mathbf{E}^{Q_t}[(
r_{d}(t) K - r_{f}(t) \ S(t)) \textbf{1}_{ \{S(t)>K\} }\ ]$ at
this point $t_1$. This allows us to calculate from equation
(\ref{locvol C}) the function $\sigma(t_1,S(t_1))$ at this time
step. Following this procedure through time, we generate
simultaneously both the expectation in (\ref{locvol C}) and the
local volatility function $\sigma(t,S(t))$. At each time step we
generate also all domestic forward joint transition densities
$\phi_F(S(t),r_d(t), r_f(t),t) $ by solving the forward PDE
(\ref{PDE phiTF}).

\subsection{Comparison between local volatility with and without stochastic interest rates \label{section: calibration Comparison between local volatility}}

In this section we present a way to calibrate the extension of the
local volatility function derived in section \ref{section :
Dupire's like formula } in a three-factor framework by using the
local volatility showing up in the simple one-factor
Gaussian model.\\

Assuming deterministic interest rates, the FX spot follows the
diffusion equation

\begin{eqnarray}
dS(t) = (r_{d}(t) - r_{f}(t)) S(t) dt + \sigma_{1f}(t,S(t))  S(t)
 dW^{DRN}_{S}(t)
\end{eqnarray}

\noindent where the local volatility function denoted by
$\sigma_{1f}$ is given by equation (\ref{detreministic locvol
C})\footnote{in the case of deterministic interest rates
$f_{d}(0,t) = r_d(t)$ and $f_{f}(0,t) = r_f(t)$}

\begin{eqnarray}
\sigma_{1f}^{2}(t,K)  = \frac{\frac{\partial C(K,t)}{\partial t} +
K (f_{d}(0,t)-f_{f}(0,t))
 \frac{\partial C(K,t)}{\partial K} + f_{f}(0,t)C(K,t) }{ \frac{1}{2}  K^{2} \frac{\partial^{2}
C(K,t)}{\partial K^{2}}}. \label{detreministic locvol C 02}
\end{eqnarray}

However, if we consider the three-factor model with stochastic
interest rates, the local volatility function is given by equation
(\ref{locvol C})

\begin{eqnarray}
\sigma_{3f}^{2}(t,K) &=& \frac{\frac{\partial C(K,t)}{\partial t}
- P_{d}(0,t) \mathbf{E}^{Q_t}[ (K r_{d}(t) - r_{f}(t) S(t))
\textbf{1}_{ \{S(t)>K\}} ] }{ \frac{1}{2}  K^{2}
\frac{\partial^{2} C(K,t)}{\partial K^{2}}}. \label{locvol C 02}
\end{eqnarray}

Using the fact that under the $T$-forward measure for $T \geq t$,
we have: $\mathbf{E}^{Q_T}[ r_{d}(T) \mid \mathcal{F}_{t} ] =
f_{d}(t,T)$ and $\mathbf{E}^{Q_T}[ r_{f}(T) \mid \mathcal{F}_{t} ]
= f_{f}(t,T)$ we notice that

\begin{eqnarray}
\mathbf{E}^{Q_t}[ r_{d}(t) \textbf{1}_{ \{S(t)>K\}} ] &=&
\mathbf{E}^{Q_t}[ r_{d}(t) ] \mathbf{E}^{Q_t}[ \textbf{1}_{
\{S(t)>K\}} ] + \mathbf{Cov}^{Q_t}[ r_{d}(t) , \textbf{1}_{
\{S(t)>K\}} ] \nonumber \\
&=& f_{d}(0,t) (- \frac{1}{P_{d}(0,t)} \frac{\partial
C(t,K)}{\partial K}) + \mathbf{Cov}^{Q_t}[ r_{d}(t) , \textbf{1}_{
\{S(t)>K\}} ] \label{resultat03}
\end{eqnarray}

\noindent where $\mathbf{Cov}^{Q_t}(X,Y)$ represents the
covariance between two stochastic variables X and Y with dynamics
expressed in the $t$-forward measure $Q_t$. We also have

\begin{eqnarray}
\mathbf{E}^{Q_t}[r_{f}(t) S(t) \textbf{1}_{ \{S(t)>K\}} ] &=&
\mathbf{E}^{Q_t}[r_{f}(t) (S(t) - K) \textbf{1}_{ \{S(t)>K\}} ]
+ \mathbf{E}^{Q_t}[r_{f}(t)  K \textbf{1}_{ \{S(t)>K\}} ] \nonumber \\
&=& \mathbf{E}^{Q_t}[r_{f}(t)] \mathbf{E}^{Q_t}[(S(t) - K)^{+} ]
+ \mathbf{Cov}^{Q_t}[r_{f}(t),(S(t) - K)^{+} ] \nonumber \\
& & + K ( \mathbf{E}^{Q_t}[r_{f}(t)]\mathbf{E}^{Q_t}[\textbf{1}_{
\{S(t)>K\}} ] +\mathbf{Cov}^{Q_t}[r_{f}(t) , \textbf{1}_{
\{S(t)>K\}} ])\nonumber \\
&=& f_{f}(0,t) \frac{C(t,K)}{P_{d}(0,t)} + \mathbf{Cov}^{Q_t}[r_{f}(t),(S(t) - K)^{+} ] \nonumber \\
& & +  K ( f_{f}(0,t) (- \frac{1}{P_{d}(0,t)} \frac{\partial
C(t,K)}{\partial K}) + \mathbf{Cov}^{Q_t}[r_{f}(t) , \textbf{1}_{
\{S(t)>K\}} ]).  \nonumber \\ \label{resultat04}
\end{eqnarray}

Substituting expressions (\ref{resultat03}) and (\ref{resultat04})
in equation (\ref{locvol C 02}), one finds the following
interesting  relation between the simple Dupire formula
(\ref{detreministic locvol C 02}) and its extension (\ref{locvol C
02})

\begin{eqnarray}
\sigma_{3f}^{2}(t,K) - \sigma_{1f}^{2}(t,K)  = \frac{ K P_{d}(0,t)
\{\mathbf{Cov}^{Q_t}[r_{f}(t) - r_{d}(t) , \textbf{1}_{
\{S(t)>K\}} ]  + \frac{1}{ K} \mathbf{Cov}^{Q_t}[r_{f}(t),(S(t) -
K)^{+} ] \} }{ \frac{1}{2} K^{2} \frac{\partial^{2} C}{\partial
K^{2}}}. \nonumber \\   \label{loc vol difference}
\end{eqnarray}

Assuming that quantities $\mathbf{Cov}^{Q_t}[r_{f}(t),(S(t) -
K)^{+} ]$ and $\mathbf{Cov}^{Q_t}[r_{f}(t) - r_{d}(t) ,
\textbf{1}_{ \{S(t)>K\}} ]$  are extractable from the market,
equation (\ref{loc vol difference}) shows the corrections to make
to the tractable Dupire local volatility surface in order to
obtain the local volatility surface which takes into account the
effects of both domestic and foreign stochastic interest rates.

\subsection{Calibrating the local volatility by mimicking stochastic volatility
models \label{section : mimicking stoch vol}}

In this section we consider the three factor model with a
stochastic volatility for the spot FX rate and we show how to
connect this model to the one where the spot FX rate volatility is
a local volatility. More precisely, we sketch the derivation of an
expectation relationship between local volatilities and future
instantaneous spot FX rate volatilities.\\

Consider the following domestic risk neutral dynamics for the spot
FX rate

\begin{eqnarray}
dS(t) = (r_{d}(t) - r_{f}(t)) S(t) dt + \gamma(t,\nu(t)) S(t)
dW^{DRN}_{S}(t) \label{DRN dynamics with stoch vol}
\end{eqnarray}

\noindent where $\nu(t)$ is a stochastic variable which provides
the stochastic perturbation for the spot FX rate volatility.
Common choices for the function $\gamma(t,\nu(t))$ are $\nu(t)$, $
exp(\sqrt{\nu(t)})$ and  $\sqrt{\nu(t)}$ . The stochastic variable
$\nu(t)$ is generally modelled by  a Cox-Ingersoll-Ross (CIR)
process (mean-reversion in the drift and a volatility dependent of
$\sqrt{\nu(t)}$ as for example the Heston model \citep{heston}) or
in the first case by a Ornstein-Uhlenbeck process (OU)
(mean-reversion in the drift and a volatility independent from
$\nu(t)$ as for example the Sch\"{o}bel and Zhu
\citep{SchobelandZhu} stochastic volatility model) which allows
for negative values of $\nu(t)$ .

Applying Tanaka's formula to the convex but non-differentiable
function $e^{- \int_{0}^{t} r_{d}(s) ds}(S(t)-K)^{+}$ leads to

\begin{eqnarray}
e^{- \int_{0}^{t} r_{d}(s)ds}(S(t)-K)^{+} &=& (S(0)-K)^{+} -
\int_{0}^{t} r_{d}(u) e^{- \int_{0}^{u} r_{d}(s)ds}(S(u)-K)^{+}
du \nonumber \\
& &  + \int_{0}^{t} e^{- \int_{0}^{u} r_{d}(s)ds} \textbf{1}_{
\{S(u)>K\} } dS_{u} + \frac{1}{2} \int_{0}^{t} e^{- \int_{0}^{u}
r_{d}(s)ds} dL_{u}^{K}(S) \nonumber
\end{eqnarray}

\noindent where $L_{u}^{K}(S)$ is the local time of $S$. Since $S$
is a continuous semimartingale, then almost-surely (see
\citep{Revuz-Yor})

\begin{eqnarray}
L_{t}^{K}(S) &=&   \lim_{\epsilon \downarrow 0} \frac{1}{\epsilon}
\int_{0}^{t} \textbf{1}_{ [K , K + \epsilon] }(S(s)) d<S,S>_s.
\label{local time}
\end{eqnarray}

Using the domestic risk neutral diffusion for the spot FX rate
(\ref{DRN dynamics with stoch vol}), one obtains

\begin{eqnarray}
e^{- \int_{0}^{t} r_{d}(s)ds}(S(t)-K)^{+}  &=& (S(0)-K)^{+} + K
\int_{0}^{t} r_{d}(u) e^{- \int_{0}^{u} r_{d}(s)ds} \textbf{1}_{
\{S(u)>K\} }\ du   \nonumber \\
& & - \int_{0}^{t} e^{- \int_{0}^{u} r_{d}(s)ds} \textbf{1}_{
\{S(u)>K\} } r_{f}(u)S(u) du \nonumber \\
& & + \int_{0}^{t} e^{- \int_{0}^{u} r_{d}(s)ds} \textbf{1}_{
\{S(u)>K\} } \gamma(u,\nu(u)) S(u)
dW^{DRN}_{S}(u)\nonumber \\
& & + \frac{1}{2} \int_{0}^{t} e^{- \int_{0}^{u} r_{d}(s)ds}
dL_{u}^{K}(S). \nonumber
\end{eqnarray}

Assuming that the function $e^{- \int_{0}^{u} r_{d}(s)ds}
\textbf{1}_{ \{S(u)>K\} } \gamma(u,\nu(u)) S(u)$ is a member of
the class $\mathcal{H}^{2}$, namely the measurable and adapted
functions $f$ such that $\mathbf{E}^{Q_{d}}[\int_{0}^{t} f^2(s)
ds] < \infty $, we get when taking the domestic risk neutral
expectation of each side

\begin{eqnarray}
\mathbf{E}^{Q_{d}}[e^{- \int_{0}^{t} r_{d}(s)ds}(S(t)-K)^{+}] &=&
\mathbf{E}^{Q_{d}}[(S(0)-K)^{+}] + K \int_{0}^{t}
\mathbf{E}^{Q_{d}}[ r_{d}(u)
e^{- \int_{0}^{u} r_{d}(s)ds} \textbf{1}_{ \{S(u)>K\} }\ ]du  \nonumber \\
& & -  \int_{0}^{t} \mathbf{E}^{Q_{d}}[ e^{- \int_{0}^{u}
r_{d}(s)ds} \textbf{1}_{ \{S(u)>K\} } r_{f}(u) \ S(u)] du \nonumber \\
& & +   \int_{0}^{t} \frac{1}{2}  \mathbf{E}^{Q_{d}}[ e^{-
\int_{0}^{u} r_{d}(s)ds} dL_{u}^{K}(S)]. \nonumber
\end{eqnarray}

Differentiating this equation leads to

\begin{eqnarray}
dC(K,t) &=&   K \mathbf{E}^{Q_{d}}[ r_{d}(t) e^{- \int_{0}^{t}
r_{d}(s)ds} \textbf{1}_{ \{S(t)>K\} }\ ] dt - \mathbf{E}^{Q_{d}}[
e^{- \int_{0}^{t} r_{d}(s)ds} \textbf{1}_{
\{S(t)>K\} } r_{f}(t) \ S(t) ] dt \nonumber \\
& &  + \frac{1}{2} \mathbf{E}^{Q_{d}}[ e^{- \int_{0}^{t}
r_{d}(s)ds} dL_{t}^{K}(S)]. \nonumber
\end{eqnarray}

Using characterization (\ref{local time}) for the local time with
$d<S,S>_t  = \gamma^{2}(t,\nu(t)) \ S^{2}(t) dt$, we obtain

\begin{eqnarray}
dC(K,t) &=&   K \mathbf{E}^{Q_{d}}[ r_{d}(t) e^{- \int_{0}^{t}
r_{d}(s)ds} \textbf{1}_{ \{S(t)>K\} }\ ] dt - \mathbf{E}^{Q_{d}}[
e^{- \int_{0}^{t} r_{d}(s)ds} \textbf{1}_{ \{S(t)>K\}
} r_{f}(t) \ S(t) ] dt  \nonumber \\
& &  + \frac{1}{2} \lim_{\epsilon  \downarrow 0}
 \mathbf{E}^{Q_{d}}[ \frac{1}{\epsilon} \textbf{1}_{ [K , K + \epsilon] }(S(t)) e^{- \int_{0}^{t}
r_{d}(s)ds} \gamma^2(t,\nu(t))  S^{2}(t) ]  dt. \label{resultat05}
\end{eqnarray}

Here, the last terms of the equation (\ref{resultat05}) can be
rewritten as

\begin{eqnarray}
& & \lim_{\epsilon  \downarrow 0} \frac{1}{\epsilon}
\mathbf{E}^{Q_{d}}[\textbf{1}_{ [K , K + \epsilon] }(S(t)) e^{-
\int_{0}^{t} r_{d}(s)ds} \gamma^2(t,\nu(t))  S^{2}(t) ] \nonumber \\
& & = \lim_{\epsilon \downarrow 0} \frac{1}{\epsilon}
\mathbf{E}^{Q_{d}}[ \mathbf{E}^{Q_{d}}[ \gamma^2(t,\nu(t)) e^{-
\int_{0}^{t} r_{d}(s)ds} \mid S(t)
] \textbf{1}_{ [K , K + \epsilon] }(S(t))   S^{2}(t) ] \nonumber \\
& &  = \mathbf{E}^{Q_{d}}[ \gamma^2(t,\nu(t)) e^{- \int_{0}^{t}
r_{d}(s)ds} \mid S(t) =
K ] p_{d}(K,t)  K^{2}  \nonumber \\
& & =  \frac{ \mathbf{E}^{Q_{d}}[\gamma^2(t,\nu(t)) e^{-
\int_{0}^{t} r_{d}(s)ds} \mid S(t) = K ] }{\mathbf{E}^{Q_{d}}[
e^{- \int_{0}^{t} r_{d}(s)ds} \mid S(t) = K ]} \frac{\partial^{2}
C(K,t)}{\partial K^{2}}  K^{2}. \label{resultat06}
\end{eqnarray}

Substituting equation (\ref{resultat06}) in (\ref{resultat05}),
one obtains

\begin{eqnarray}
\frac{ \mathbf{E}^{Q_{d}}[\gamma^2(t,\nu(t)) e^{- \int_{0}^{t}
r_{d}(s)ds} \mid S(t) = K ] }{\mathbf{E}^{Q_{d}}[ e^{-
\int_{0}^{t} r_{d}(s)ds} \mid S(t) = K ]} = \frac{\frac{\partial
C}{\partial t} - \mathbf{E}^{Q_d}[ e^{- \int_{0}^{t} r_{d}(s)ds}
(K r_{d}(t) - r_{f}(t) S(t)) \textbf{1}_{ \{S(t)>K\}} ] }{
\frac{1}{2}  K^{2} \frac{\partial^{2} C}{\partial K^{2}}}.
\label{SV with C}
\end{eqnarray}

The right hand side of this equation equals the local volatility
in the three-factor model where the expectation is expressed under
the domestic risk-neutral measure. Therefore, if there exists a
local volatility such that the one-dimensional probability
distribution of the spot FX rate with the diffusion
(\ref{FXspot_dynamic}) is the same as the one of the spot FX rate
with dynamics (\ref{DRN dynamics with stoch vol}) for every time
$t$, then this local volatility function has to satisfy the
following equation

\begin{eqnarray}
\sigma^{2}(t,K)  =   \frac{ \mathbf{E}^{Q_{d}}[\gamma^2(t,\nu(t))
e^{- \int_{0}^{t} r_{d}(s)ds} \mid S(t) = K ]
}{\mathbf{E}^{Q_{d}}[ e^{- \int_{0}^{t} r_{d}(s)ds} \mid S(t) = K
]}. \label{resultat07}
\end{eqnarray}

Stochastic volatility models belong to the class of incomplete
market models, and hence, we have to precise that this condition
holds if we assume that the domestic risk neutral probability
measure $Q_d$ used in both stochastic and local volatility
framework are the same. This equation extends the result obtained
by Dupire \citep{Dupire-1998} and Derman and Kani
\citep{Derman1998} to the case where domestic and foreign interest
rates are stochastic. We notice that when we assume deterministic
interest rates, equation (\ref{resultat07}) reduces exactly to
their result

\begin{eqnarray}
\sigma^{2}(t,K)  = \mathbf{E}^{Q_{d}}[\gamma^2(t,\nu(t))  \mid
S(t) = K ].
\end{eqnarray}

Finally, we reduce equation (\ref{resultat07}) to a conditional
expectation under the $t$-forward measure

\begin{eqnarray}
\sigma^{2}(t,K) =  \mathbf{E}^{Q_t}[ \gamma^2(t,\nu(t)) \mid S(t)
= K ]. \label{expectation for calibration}
\end{eqnarray}

The local volatility $\sigma^{2}(t,S(t)=K)$ is, therefore, the
conditional expectation under the $t$-forward measure of the
instantaneous spot FX rate volatility at the future time $t$,
contingent on the spot FX rate level $S(t)$ being equal to $K$.\\



Equation (\ref{expectation for calibration}) give us a new way to
calibrate the local volatility function. A numerical approach is
to begin with solving the forward PDE (\ref{PDE phiTF}) by using
for example finite differences or finite elements methods, to
determine the forward probability density $\phi_{F}$ and
afterwards using numerical integration methods in order to
calculate the conditional expectation (\ref{expectation for
calibration}).\\

Conditional expectations are difficult to compute by using
traditional Monte Carlo simulations since paths generated by the
simulation will miss the event involved in the conditional
expectation. However, Malliavin integration by parts allows to
obtain different representations of such conditional expectation
that can be compute efficiently by Monte Carlo simulations (see
\citep{Fournie2}).


\begin{Rem}
If we assume independence between the spot FX rate and its
volatility, the local volatility function is given by
$\sigma^{2}(T,K) =  \mathbf{E}^{Q_T}[ \gamma^2(T,\nu(T))]$. In
some particular cases it is possible to derive closed-form
solutions for this expectation. Consider for example the
three-factor model with local volatility where domestic and
foreign interest rates have Hull and White dynamics (see section
\ref{section : the model}) and we want to find the local
volatility function by using the mimicking property
$(\ref{expectation for calibration})$ derived from a four-factor
model with the same interest rates dynamics and with Sch\"{o}bel
and Zhu dynamics for the FX spot volatility ($\nu(t)$),

\begin{numcases}{}
dS(t) = (r_{d}(t) - r_{f}(t))  S(t)  dt + \nu(t)  S(t)  dW^{DRN}_{S}(t), &  \\
dr_{d}(t) = [\theta_{d}(t) - \alpha_{d}  r_{d}(t) ]  dt +
\sigma_{d} dW^{DRN}_{d}(t), & \\
dr_{f}(t) = [\theta_{f}(t) - \alpha_{f} r_{f}(t) - \rho_{fS}
\sigma_{f} \nu(t) ]  dt + \sigma_{f}
dW^{DRN}_{f}(t),&  \\
d \nu(t) = k [\lambda-\nu(t)] \ dt + \xi  dW^{DRN}_{\nu}(t),
\label{Schobel and Zhu dynamics} &
\end{numcases}

\noindent Under the $T$-Forward measure, the Sch\"{o}bel and Zhu
dynamics $(\ref{Schobel and Zhu dynamics})$ become

\begin{eqnarray}
 d \nu(t) = k (\Lambda(t) -\nu(t))  dt + \xi \ dW^{TF}_{\nu}(t),
 \label{Schobel_Zhu_Tfwd}
\end{eqnarray}

\noindent where $\Lambda(t) = \lambda - \frac{\rho_{d \nu}
\sigma_d b_d(t,T) \xi}{k}$, $b_d(t,T)= \frac{1}{\alpha_{d}}
(1-e^{-\alpha_{d}(T-t)})$\\

Integrating equation $(\ref{Schobel_Zhu_Tfwd})$, we obtain, for
each $t\leq T$

\begin{eqnarray}
\nu(T) = \nu(t) e^{-k(T-t)}  + \int_{t}^{T} k \Lambda(u)
e^{-k(T-u)} du  + \int_{t}^{T}  \xi e^{-k(T-t)} dW^{TF}_{\nu}(u),
\end{eqnarray}

\noindent so that $\nu(T)$ conditional on $\mathcal{F}_t$ is
normally distributed with mean and variance given respectively by

\begin{eqnarray}
\mathbf{E}^{Q_{T}}[\nu(T)| \mathcal{F}_t ] &=& \nu(t) e^{-k(T-t)}
+ (\lambda - \frac{\rho_{d \nu} \sigma_d \xi}{\alpha_{d}k}) (1 -
e^{-k(T-t)}) + \frac{\rho_{d \nu} \sigma_d \xi}{\alpha_{d}
(\alpha_{d}+k)) } (1- e^{-(\alpha_{d}+k)(T-t)}) \nonumber \\
\mathbf{Var}^{Q_{T}}[\nu(T) |\mathcal{F}_t ] &=& \frac{\xi^2}{2k}
(1- e^{-2k(T-t)}) \nonumber
\end{eqnarray}

Finally, the closed form solution for the local volatility
function is given by

\begin{eqnarray}
\sigma^{2}(T,K) &=& \mathbf{E}^{Q_{T}}[\nu^2(T)] =
(\mathbf{E}^{Q_{T}}[\nu(T)])^2 + \mathbf{Var}^{Q_{T}}[\nu(T) ]
\nonumber \\
&=& \left( \nu(t) e^{-kT} + (\lambda - \frac{\rho_{d \nu} \sigma_d
\xi}{\alpha_{d}k}) (1 - e^{-kT}) + \frac{\rho_{d \nu} \sigma_d
\xi}{\alpha_{d} (\alpha_{d}+k)) } (1- e^{-(\alpha_{d}+k)T})
\right)^2 + \frac{\xi^2}{2k} (1- e^{-2kT}) \nonumber \\
& &
\end{eqnarray}

\end{Rem}

\section{Hybrid volatility model \label{section:Extension}}

In a simple local volatility model, the instantaneous volatility
of the spot FX rate is a deterministic function of time and spot
FX level. As a consequence, these models are suitable for pricing
derivatives in situations where the spot FX volatility is strongly
correlated to the spot FX market level itself. However, in the FX
option's market, the volatility seems to exhibit some stochastic
behavior especially in the long dated market. In this section we
study an extension of the three-factor model with local volatility
which incorporates stochastic behavior in the spot FX volatility
without deleting the local volatility one.\\

In this section we consider a hybrid volatility model where the
volatility of the spot FX rate is a combination of a stochastic
and a local volatility function. More precisely, the volatility
for the spot FX rate corresponds to a local volatility
$\sigma_{LOC2}(t,S(t))$ multiplied by a stochastic volatility
$\gamma(t,\nu(t))$ where $\nu(t)$ is a stochastic variable. This
gives the following four-factor model with local volatility

\begin{numcases}{}
dS(t) = (r_{d}(t) - r_{f}(t)) S(t) dt + \sigma_{LOC2}(t,S(t)) \gamma(t,\nu(t)) S(t) dW^{DRN}_{S}(t), & \label{hybrid vol mod} \\
dr_{d}(t) = [\theta_{d}(t) - \alpha_{d}(t)  r_{d}(t) ]  dt + \sigma_{d}(t) dW^{DRN}_{d}(t), & \nonumber \\
dr_{f}(t) = [\theta_{f}(t) - \alpha_{f}(t)  r_{f}(t) - \rho_{fS}
\sigma_{f}(t) \sigma_{LOC2}(t,S(t)) \gamma(t,\nu(t)) ]  dt +
\sigma_{f}(t)
dW^{DRN}_{f}(t),\nonumber & \\
d\nu(t) = \alpha(t,\nu(t)) dt + \vartheta(t,\nu(t))
dW^{DRN}_{\nu}(t). \nonumber &
\end{numcases}

We will show how to connect this model to the one where the spot FX
rate volatility is a pure local volatility. In this case the
volatility of the spot FX rate is modelled by a local volatility
denoted by $\sigma_{LOC1}(t,S(t))$

\begin{numcases}{}
dS(t) = (r_{d}(t) - r_{f}(t)) S(t) dt + \sigma_{LOC1}(t,S(t)) S(t) dW^{DRN}_{S}(t), & \nonumber \\
dr_{d}(t) = [\theta_{d}(t) - \alpha_{d}(t)  r_{d}(t) ]  dt + \sigma_{d}(t) dW^{DRN}_{d}(t), & \nonumber \\
dr_{f}(t) = [\theta_{f}(t) - \alpha_{f}(t)  r_{f}(t) - \rho_{fS}
\sigma_{f}(t) \sigma_{LOC1}(t,S(t)) ]  dt + \sigma_{f}(t)
dW^{DRN}_{f}(t).\label{pure loc vol mod} &
\end{numcases}

Consider the domestic risk neutral dynamics for the spot FX rate
in the Hybrid volatility model (equation(\ref{hybrid vol mod}))

\begin{eqnarray}
dS(t) = (r_{d}(t) - r_{f}(t)) S(t) dt + \sigma_{LOC2}(t,S(t))
\gamma(t,\nu(t)) S(t) dW^{DRN}_{S}(t) \label{DRN dynamics with
stoch vol and loc vol}
\end{eqnarray}

\noindent where $\nu(t)$ is a stochastic variable which provides
the stochastic perturbation for the spot FX rate volatility.
Applying Tanaka's formula to the non-differentiable function $e^{-
\int_{0}^{t} r_{d}(s) ds} \ (S(t)-K)^{+}$ and following the same
steps as in section \ref{section : mimicking stoch vol} leads to

\begin{eqnarray}
dC(K,t) &=&   K \mathbf{E}^{Q_{d}}[ r_{d}(t) e^{- \int_{0}^{t}
r_{d}(s)ds} \textbf{1}_{ \{S(t)>K\} }\ ] dt - \mathbf{E}^{Q_{d}}[
e^{- \int_{0}^{t} r_{d}(s)ds} \textbf{1}_{ \{S(t)>K\}
} r_{f}(t)  S(t) ] dt  \nonumber \\
& &  + \frac{1}{2} \lim_{\epsilon  \downarrow 0}
 \mathbf{E}^{Q_{d}}[ \frac{1}{\epsilon} \textbf{1}_{ [K , K + \epsilon] }(S(t)) e^{- \int_{0}^{t}
r_{d}(s)ds} \sigma^2_{LOC2}(t,S(t)) \gamma^2(t,\nu(t))  S^{2}(t) ]
dt. \label{resultat05}
\end{eqnarray}

Here, the last terms of the equation (\ref{resultat05}) can be
rewritten as

\begin{eqnarray}
& & \lim_{\epsilon  \downarrow 0} \frac{1}{\epsilon}
\mathbf{E}^{Q_{d}}[\textbf{1}_{ [K , K + \epsilon] }(S(t)) e^{-
\int_{0}^{t} r_{d}(s)ds} \sigma^2_{LOC2}(t,S(t)) \gamma^2(t,\nu(t))  S^{2}(t) ] \nonumber \\
& & = \lim_{\epsilon \downarrow 0} \frac{1}{\epsilon}
\mathbf{E}^{Q_{d}}[ \mathbf{E}^{Q_{d}}[ \gamma^2(t,\nu(t)) e^{-
\int_{0}^{t} r_{d}(s)ds} \mid S(t)
] \textbf{1}_{ [K , K + \epsilon] }(S(t))   \sigma^2_{LOC2}(t,S(t)) S^{2}(t) ] \nonumber \\
& &  = \mathbf{E}^{Q_{d}}[ \gamma^2(t,\nu(t)) e^{- \int_{0}^{t}
r_{d}(s)ds} \mid S(t) =
K ]  p_{d}(K,t) \sigma^2_{LOC2}(t,K)  K^{2}  \nonumber \\
& & =  \frac{ \mathbf{E}^{Q_{d}}[\gamma^2(t,\nu(t)) e^{-
\int_{0}^{t} r_{d}(s)ds} \mid S(t) = K ] }{\mathbf{E}^{Q_{d}}[
e^{- \int_{0}^{t} r_{d}(s)ds} \mid S(t) = K ]}
 \frac{\partial^{2} C(K,t)}{\partial K^{2}}
\sigma^2_{LOC2}(t,K) K^{2}. \label{resultat06}
\end{eqnarray}

Substituting equation (\ref{resultat06}) in (\ref{resultat05}),
one obtains

\begin{eqnarray}
\sigma^2_{LOC2}(t,K) \frac{ \mathbf{E}^{Q_{d}}[\gamma^2(t,\nu(t))
e^{- \int_{0}^{t} r_{d}(s)ds} \mid S(t) = K ]
}{\mathbf{E}^{Q_{d}}[ e^{- \int_{0}^{t} r_{d}(s)ds} \mid S(t) = K
]} = \frac{\frac{\partial C}{\partial t} - \mathbf{E}^{Q_d}[ e^{-
\int_{0}^{t} r_{d}(s)ds} (K r_{d}(t) - r_{f}(t) S(t)) \textbf{1}_{
\{S(t)>K\}} ] }{ \frac{1}{2}  K^{2} \frac{\partial^{2} C}{\partial
K^{2}}}. \nonumber \\ \label{SV with C}
\end{eqnarray}

The right hand side of this equation equals the local volatility
in the three-factor model where the expectation is expressed under
the domestic risk-neutral measure. Therefore, if there exists a
local volatility such that the one-dimensional probability
distribution of the spot FX rate with the diffusion (\ref{pure loc
vol mod}) is the same as the one of the spot FX rate with dynamics
(\ref{DRN dynamics with stoch vol and loc vol}) for every time
$t$, then this local volatility function has to satisfy the
following equation

\begin{eqnarray}
\sigma^{2}_{LOC1}(t,K)  &=& \sigma^2_{LOC2}(t,K)  \frac{
\mathbf{E}^{Q_{d}}[\gamma^2(t,\nu(t)) e^{- \int_{0}^{t}
r_{d}(s)ds} \mid S(t) = K ] }{\mathbf{E}^{Q_{d}}[ e^{-
\int_{0}^{t} r_{d}(s)ds} \mid S(t) = K ]} \label{resultat07} \\
&=& \sigma^2_{LOC2}(t,K) \mathbf{E}^{Q_{t}}[\gamma^2(t,\nu(t))
\mid S(t) = K ]. \label{resultat08}
\end{eqnarray}

Hybrid volatility models belong also to the class of incomplete
market models, and hence, we assume that the domestic risk neutral
probability measure $Q_d$ used in both hybrid and pure local
volatility framework are the same. This equation extends the
result obtained by Madan \citep{Madan} to the case where domestic
and foreign interest rates are stochastic. We notice that when we
assume deterministic interest rates, equation (\ref{resultat07})
reduces exactly to their result

\begin{eqnarray}
\sigma^2_{LOC1}(t,K)  = \sigma^2_{LOC2}(t,K)
\mathbf{E}^{Q_{d}}[\gamma^2(t,\nu(t))  \mid S(t) = K ].
\end{eqnarray}

The local volatility $\sigma^2_{LOC2}(t,K)$ is, therefore, given
by the local volatility $\sigma^2_{LOC1}(t,K)$ divided by the
conditional expectation under the $t$-forward measure of the
instantaneous spot FX rate volatility at the future time $t$,
contingent on the spot FX rate level $S(t)$ being equal to $K$.\\

\begin{eqnarray}
\sigma^2_{LOC2}(t,K)  =
\frac{\sigma^2_{LOC1}(t,K)}{\mathbf{E}^{Q_t}[ \gamma^2(t,\nu(t))
\mid S(t) = K ]}. \label{hybrid expectation for calibration}
\end{eqnarray}

Equation (\ref{hybrid expectation for calibration}) give us a way to
calibrate the local volatility function $\sigma_{LOC2}(t,K)$
knowing the local volatility function $\sigma_{LOC1}(t,K)$. A
numerical approach is to begin with solving the forward PDE
(\ref{PDE phiTF}) by using for example finite differences or
finite elements methods, to determine the forward probability
density $\phi_{F}$ and afterwards using numerical integration
methods in order to calculate the conditional expectation
(\ref{hybrid expectation for calibration}) (see subsection
\ref{calibration hybrid}). As mentioned in section \ref{section :
mimicking stoch vol}, an other approach is to use Malliavin
integration by parts to obtain different representations of the
conditional expectation that can be computed efficiently by Monte
Carlo simulations (see
\citep{Fournie2}).\\

\begin{Rem}
If we assume independence between the spot FX rate and its
volatility, the local volatility function is given by
$\sigma^2_{LOC2}(t,K)  =
\frac{\sigma^2_{LOC1}(t,K)}{\mathbf{E}^{Q_t}[
\gamma^2(t,\nu(t))]}$. Consider the hybrid model with Sch\"{o}bel
and Zhu dynamics for the stochastic component of the FX spot
volatility ($\nu(t)$),

\begin{numcases}{}
dS(t) = (r_{d}(t) - r_{f}(t))  S(t)  dt + \sigma_{LOC2}(t,S(t)) \nu(t)  S(t)  dW^{DRN}_{S}(t), &  \\
dr_{d}(t) = [\theta_{d}(t) - \alpha_{d}  r_{d}(t) ]  dt +
\sigma_{d} dW^{DRN}_{d}(t), & \\
dr_{f}(t) = [\theta_{f}(t) - \alpha_{f} r_{f}(t) - \rho_{fS}
\sigma_{f} \nu(t) ]  dt + \sigma_{f}
dW^{DRN}_{f}(t),&  \\
d \nu(t) = k [\lambda-\nu(t)] \ dt + \xi  dW^{DRN}_{\nu}(t),
\label{Schobel and Zhu dynamics hybrid} &
\end{numcases}

In this particular case we have a closed form solution for the
local volatility function $\sigma_{LOC2}$, (knowing the local
volatility function $\sigma_{LOC1}$) given by

\begin{eqnarray}
\sigma^2_{LOC2}(t,K)  = \frac{\sigma^2_{LOC1}(t,K)}{\left( \nu(t)
e^{-kT} + (\lambda - \frac{\rho_{d \nu} \sigma_d
\xi}{\alpha_{d}k}) (1 - e^{-kT}) + \frac{\rho_{d \nu} \sigma_d
\xi}{\alpha_{d} (\alpha_{d}+k)) } (1- e^{-(\alpha_{d}+k)T})
\right)^2 + \frac{\xi^2}{2k} (1- e^{-2kT})}. \label{expectation
for calibration}
\end{eqnarray}

\end{Rem}

In this section we obtained a link between the pure local
volatility model (\ref{pure loc vol mod}) and the hybrid
volatility model (\ref{hybrid vol mod}) such that the
one-dimensional marginal distributions of $S_t$ is the same for
every $t$ in both models. Using this result for the calibration of
the local volatility $\sigma_{LOC2}(t,S(t))$ (see section
\ref{calibration hybrid}) will guarantee that prices of Vanilla
options will be the same in both models since these products are
fully determined by the conditional probability of the spot FX
rate at the maturity time $T$ conditional to the actual spot FX
rate value. However, when we price exotic and path dependent
options where payoff depends on intermediate spot values, pricing
requires the product of all conditional probabilities. By using
this link, we have never imposed that these probabilities are the
same. As a consequence, prices coming from the hybrid volatility
model (\ref{hybrid vol mod}) are different from those coming from
the pure local volatility one (\ref{pure loc vol mod}) as well as
the pure stochastic volatility one (\ref{DRN dynamics with stoch
vol}) in the case of exotic derivatives. The advantage of the
hybrid volatility model is that we have a combination of a local
and a stochastic volatility effect in the volatility of the spot
FX rate which is more realistic. Consequently, hybrid volatility
models are generally more consistent with the exotic option's
market after being calibrated with respect to the vanilla market
(see \citep{lipton3}).

\subsection{Calibration \label{calibration hybrid}}

In this subsection we are interested in the calibration of the
local volatility $\sigma_{LOC2}(t,S(t))$ of the four-factor hybrid
volatility model knowing the local volatility
$\sigma_{LOC1}(t,S(t))$ associated to the three-factor model that
we have studied in the first part of this paper. In \citep{Madan},
the authors obtain a fast and exact calibration of vanilla options
in the case of equity options with deterministic interest rates.
Based on results obtained in this section, we extend this
calibration procedure to the case
of FX derivatives in the settings of stochastic interest rates.  \\

Remember that parameters present in the Hull-White one-factor
dynamics for the domestic and foreign interest rates,
$\theta_{d}(t), \alpha_{d}(t), \sigma_{d}(t), \theta_{f}(t),
\alpha_{f}(t), \sigma_{f}(t)$, are chosen to match European
swaption / cap-floors values in their respective currencies. The
four correlation coefficients of the model, $\rho_{Sd},
\rho_{Sf}$, $\rho_{df}$ and $\rho_{S \nu}$ are deduced from
historical data. In this hybrid volatility model one also has to
calibrate the process for $\nu(t)$. The volatility of the variance
$\vartheta(t,\nu(t))$ can be deduced from historical time series
with for example daily realized volatilities. If one uses a CIR
process to model the stochastic variable $\vartheta(t,\nu(t))$,
then it is of the form $\vartheta(t,\nu(t))=\beta(t)
\sqrt{\nu(t)}$, while if one models it by using an OU process,
$\vartheta(t,\nu(t))$ is independent of $\nu(t)$ and given by
$\vartheta(t,\nu(t)) = \beta(t)$. Usually, practitioners use a
mean-reverting form for the drift of the variance,
$\alpha(t,\nu(t))= \lambda(t)- \kappa(t) \nu(t)$. One also has to
calibrate the local volatility $\sigma_{LOC1}(t,S(t))$ associated
to the three-factor model with a local volatility only. This
calibration can be realized using methods developed in section
\ref{section : loc vol calibration}. Once all these parameters
have been determined, one is able to calibrate the local
volatility function $\sigma_{LOC2}(t,S(t))$ associated to the
hybrid volatility model by using equation (\ref{hybrid expectation
for calibration}), namely

\begin{eqnarray}
\sigma^2_{LOC2}(t,K)  =
\frac{\sigma^2_{LOC1}(t,K)}{\mathbf{E}^{Q_t}[ \gamma^2(t,\nu(t))
\mid S(t) = K ]}, \label{equation pour calib vol loc2}
\end{eqnarray}
where the conditional expectation is by definition given by

\begin{eqnarray}
 & &\mathbf{E}^{Q_t}[ \gamma^2(t,\nu(t))
\mid S(t) = K ] \nonumber \\
 & & = \frac{\int_{0}^{\infty} \gamma^2(t,\nu(t))
\phi_F(S(t)=K, r_d(t), r_f(t),\nu(t),t) d\nu}{\int_{0}^{\infty}
\phi_F(S(t)=K,
r_d(t), r_f(t),\nu(t),t) d\nu}. \label{esp cond calibration2}  
\end{eqnarray}

The $t$-forward joint density $\phi_F(S(t),r_d(t), r_f(t),\nu(t),t)$
satisfies a four-dimensional Kolmogorov forward equation that can
be derived using the same method as in section \ref{section : fwd PDE}.\\

The strategy is to solve this forward equation forwards one step
at a time, starting with a local volatility
$\sigma_{LOC2}(0,S(0))$ equal to $\sigma_{LOC1}(0,S(0)) /
\gamma(0,\nu(0)) $ at time $t_0=0$, since the volatility $\nu(0)$
is known and we have $ \mathbf{E}^{Q_t}[
 \gamma(0,\nu(0)) | S(0)=K] = \gamma(0,\nu(0))$. At the time step
$t_1=t_0+\Delta t$ we can generate the forward joint transition
densities $ \phi_F(x,y,z,\nu,t_0+\Delta t) $ by solving the
forward PDE using the initial condition
$\phi_{F}(x,y,z,\nu,t_0)=\delta(x-x_0,y-y_0,z-z_0,\nu-\nu_0)$ at
time $t_0=0$. Knowing the $t$-forward joint transition densities we
can calculate each integral in equation (\ref{esp cond
calibration2}) to determine the function $ \mathbf{E}^{Q_t}[
 \gamma(t_1,\nu(t_1)) | S(t_1)=K]$ at this point. This allows us to
calculate from equation (\ref{equation pour calib vol loc2}) the
function $\sigma_{LOC2}(t_1,S(t_1))$ at this time step. Following
this procedure through time, we generate simultaneously both the
conditional expectation $ \mathbf{E}^{Q_t}[
 \gamma(t,\nu(t)) | S(t)=K]$ and the local volatility function
$\sigma_{LOC2}(t,S(t))$. At each time step we generate also all
forward joint transition densities $ \phi_F(S(t),r_d(t),
r_f(t),\nu(t),t) $ by solving the forward PDE.

\section{Conclusion \label{section : Conclusion}}


We have derived the local volatility expression of the spot FX
rate in a stochastic interest rates framework. Therefore this
model is very promising in pricing and hedging long-dated FX
derivatives which are more and more traded in the FX option's
market. We have proposed four different approaches for the
calibration of this local volatility function. First we have
proposed two numerical approaches based on Monte Carlo methods and
numerical resolution of a PDE, respectively. The third one is
based on the difference between the tractable local volatility
surface that exists in the context of deterministic interest rates
and our generalized one. The last method consists in generating
the local volatility surface from a stochastic volatility. More
precisely, we have obtained an explicit link between the local
volatility and the stochastic volatility and this in a stochastic
interest rates framework. This method has the advantage to give a
smoother and more stable local volatility surface than when it is
built from market available option prices.

Afterwards, we have considered an extension of the previous model
which allows the volatility of the spot FX rate to have local and
stochastic behavior. Indeed, we have studied a hybrid volatility model,
where the volatility of the spot FX rate is the product of a local
volatility and a stochastic volatility. We have obtained results
useful for the calibration of this new local volatility with
respect to the stochastic volatility and the local volatility
studied in the first part of the paper.

The three-factor model with stochastic volatility and its
extension will be useful in the pricing and risk management of
long-dated FX derivatives for which it is especially important to
consider the risk of both domestic/foreign interest rates and the
risk linked to the FX spot volatility. Future studies will include
numerical tests of the calibration methods on the FX market and
comparison between their speed and appropriateness to fit the
market implied volatility surface. The impact of stochastic
interest rates and hedging performance of these models are also
left for future research.\\

{\normalsize
\bibliographystyle{plainnat}
\bibliography{bibliography}

\begin{thebibliography}{26}
\expandafter\ifx\csname natexlab\endcsname\relax\def\natexlab#1{#1}\fi
\expandafter\ifx\csname url\endcsname\relax
  \def\url#1{{\tt #1}}\fi

\bibitem[Ahlip(2008)]{Ahlip}
R.~Ahlip.
\newblock Foreign exchange options under stochastic volatility and stochastic
  interest rates.
\newblock {\em International Journal of Theoretical and Applied Finance},
  11:\penalty0 277--294, 2008.

\bibitem[Andreasen(2006)]{Andreasen}
J.~Andreasen.
\newblock Closed form pricing of {FX} options under stochastic rates and
  volatility.
\newblock In {\em Global Derivatives Conference, ICBI}, May 2006.

\bibitem[Antonov et~al.(2008)Antonov, Arneguy, and Audet]{Antonov}
A.~Antonov, M.~Arneguy, and N.~Audet.
\newblock Markovian projection to a displaced volatility {H}eston model.
\newblock Working paper, 2008.
\newblock Available at http://ssrn.com/abstract=1106223.

\bibitem[Bossens et~al.(2010)Bossens, Rayee, Skantzos, and Deelstra]{Bossens}
F.~Bossens, G.~Rayee, N.S. Skantzos, and G.~Deelstra.
\newblock Vanna-{V}olga methods applied to {FX} derivatives: from theory to
  market practice.
\newblock {\em International Journal of Theoretical and Applied Finance},
  13(8):\penalty0 1293–1324, 2010.

\bibitem[Brigo and Mercurio(2006)]{Brigo-Mercurio}
D.~Brigo and F.~Mercurio.
\newblock {\em Interest {R}ate {M}odels - {T}heory and {P}ractice: {W}ith
  {S}mile, {I}nflation and {C}redit.}
\newblock Springer-Verlag, 2nd edition, 2006.

\bibitem[Derman and Kani(1994)]{Derman}
E.~Derman and I.~Kani.
\newblock Riding on a {S}mile.
\newblock {\em Risk}, pages 32--39, 1994.

\bibitem[Derman and Kani(1998)]{Derman1998}
E.~Derman and I.~Kani.
\newblock Stochastic implied trees: {A}rbitrage pricing with stochastic term
  and strike structure of volatility.
\newblock {\em International Journal of Theoretical and Applied Finance},
  1:\penalty0 61--110, 1998.

\bibitem[Dupire(1994)]{Dupire}
B.~Dupire.
\newblock Pricing with a {S}mile.
\newblock {\em Risk}, pages 18--20, 1994.

\bibitem[Dupire(2004)]{Dupire-1998}
B.~Dupire.
\newblock {\em A {U}nified {T}heory of {V}olatility in {D}erivatives {P}ricing:
  The {C}lassic {C}ollection}, chapter~6, pages 185--196.
\newblock (Ed. P. Carr) Risk Books, 2004.

\bibitem[Fournié et~al.(2001)Fournié, Lasry, Lebuchoux, and Lions]{Fournie2}
E.~Fournié, J.~M. Lasry, J.~Lebuchoux, and P.~L. Lions.
\newblock Applications of malliavin calculus to {M}onte-{C}arlo methods in
  finance. ii.
\newblock {\em Finance and Stochastics}, 5\penalty0 (2):\penalty0 201--236,
  2001.

\bibitem[Grzelak and Oosterlee(2010)]{Grzelak}
L.A. Grzelak and C.W. Oosterlee.
\newblock On {C}ross-{C}urrency {M}odels with {S}tochastic {V}olatility and
  {C}orrelated {I}nterest {R}ates.
\newblock Working paper, Delft University of Technology, 2010.
\newblock Available at
  http://papers.ssrn.com/sol3/papers.cfm?abstract\_id=1618684.

\bibitem[Heston(1993)]{heston}
S.L. Heston.
\newblock A closed-form solution for options with stochastic volatility with
  applications to bond and currency options.
\newblock {\em Rev Fin Studies}, 6:\penalty0 327–343, 1993.

\bibitem[Hull and White(1993)]{HullandWhite}
J.~Hull and A.~White.
\newblock One {F}actor {I}nterest {R}ate {M}odels and the {V}aluation of
  {I}nterest {R}ate {D}erivative {S}ecurities.
\newblock {\em Journal of Financial and Quantitative Analysis}, 28:\penalty0
  235–254, 1993.

\bibitem[Lipton(2002)]{lipton2}
A.~Lipton.
\newblock The vol smile problem.
\newblock {\em Risk Magazine}, 15:\penalty0 61–65, 2002.

\bibitem[Lipton and McGhee(2002)]{lipton3}
A.~Lipton and W.~McGhee.
\newblock Universal {B}arriers.
\newblock {\em Risk Magazine}, 15:\penalty0 81–85, 2002.

\bibitem[Madan et~al.(2007)Madan, Qian~Qian, and Ren]{Madan}
D.~Madan, M.~Qian~Qian, and Y.~Ren.
\newblock Calibrating and pricing with embedded local volatility models.
\newblock {\em Risk}, 2007.

\bibitem[Overhaus et~al.(2006)Overhaus, Bermudez, Buehler, Ferraris, Jordinson,
  and Lamnouar]{Overhaus}
M.~Overhaus, A.~Bermudez, H.~Buehler, A.~Ferraris, C.~Jordinson, and
  A.~Lamnouar.
\newblock {\em Equity Hybrid Derivatives}.
\newblock Wiley Finance, 2006.

\bibitem[Piterbarg(2006)]{Piterbarg2005}
V.~Piterbarg.
\newblock Smiling hybrids.
\newblock {\em Risk}, pages 66--71, 2006.

\bibitem[Revuz and Yor(2001)]{Revuz-Yor}
D.~Revuz and M.~Yor.
\newblock {\em Continuous Martingales and Brownian Motion}.
\newblock Springer-Verlag, 2001.

\bibitem[Sch\"{o}bel and Zhu(1999)]{SchobelandZhu}
R.~Sch\"{o}bel and J.~Zhu.
\newblock Stochastic volatility with an {O}rnstein {U}hlenbeck process: {A}n
  extension.
\newblock {\em European Finance Review}, 4:\penalty0 23--46, 1999.

\bibitem[Schoutens et~al.(2004)Schoutens, Simons, and Tistaert]{Schoutens}
W.~Schoutens, E.~Simons, and J.~Tistaert.
\newblock A {P}erfect {C}alibration! {N}ow {W}hat?
\newblock {\em Wilmott Magazine}, 2004.

\bibitem[Sippel and Ohkoshi(2002)]{SippelandOhkoshi}
J.~Sippel and S.~Ohkoshi.
\newblock All power to {PRDC} notes.
\newblock {\em Risk Magazine}, 15, 2002.

\bibitem[Tavella et~al.(2006)Tavella, Giese, and Vermeiren]{Tavella}
D.~Tavella, A.~Giese, and D.~Vermeiren.
\newblock Hybrid {S}tochastic {V}olatility {C}alibration.
\newblock {\em The Best of Wilmott}, 2:\penalty0 221--228, 2006.

\bibitem[van Haastrecht et~al.(2009)van Haastrecht, Lord, Pelsser, and
  Schrager]{vanHaastrecht01}
A.~van Haastrecht, R.~Lord, A.~Pelsser, and D.~Schrager.
\newblock {P}ricing {L}ong-{M}aturity {E}quity and {FX} {D}erivatives with
  {S}tochastic {I}nterest {R}ates and {S}tochastic {V}olatility.
\newblock {\em Insurance: Mathematics and Economics}, 45(3):\penalty0 436--448,
  2009.

\bibitem[van Haastrecht and Pelsser(2009)]{vanHaastrecht02}
A.~van Haastrecht and A.~Pelsser.
\newblock {G}eneric {P}ricing of {FX}, {I}nflation and {S}tock {O}ptions
  {U}nder {S}tochastic {I}nterest {R}ates and {S}tochastic {V}olatility.
\newblock Working paper (forthcoming in quantitative finance ), 2009.
\newblock Available at
  http://papers.ssrn.com/sol3/papers.cfm?abstract\_id=1197262.

\bibitem[Wilmott(2006)]{Wilmott2}
P.~Wilmott.
\newblock {\em Paul Wilmott on Quantitative Finance}.
\newblock John Wiley and Sons Ltd, 2nd edition, 2006.

\end{thebibliography}
}


\end{document}